\documentclass[12pt]{article}
\usepackage[dvips]{graphicx}
\usepackage{amsfonts}
\usepackage{amssymb,amscd,amsmath,amsthm}
\usepackage{latexsym,amstext}
\usepackage{color}
\usepackage{graphicx}

\input{tcilatex}
\begin{document}

\title{\textbf{Gamow vectors formalism applied to the Loschmidt echo}}
\author{ Sebastian Fortin \\
CONICET - Universidad de Buenos Aires, Buenos Aires, Argentina \and Manuel
Gadella \\
Departamento de F\'{\i}sicaTe\'{o}rica, At\'{o}mica y \'{O}ptica and IMUVA,\\
Universidad deValladolid, 47011Valladolid, Spain \and Federico Holik \\
Instituto de F\'{\i}sica La Plata, UNLP, CONICET, \\
Facultad de Ciencias Exactas, La Plata, Argentina \and Marcelo Losada \\
CONICET - FAMAF, Universidad Nacional de C\'{o}rdoba, C\'{o}rdoba, Argentina}
\maketitle

\begin{abstract}
Gamow vectors have been developed in order to give a mathematical
description for quantum decay phenomena. Mainly, they have been applied to
radioactive phenomena, scattering and to some decoherence models. They play
a crucial role in the description of quantum irreversible processes, and in
the formulation of time asymmetry in quantum mechanics. In this paper, we
use this formalism to describe a well-known phenomenon of irreversibility:
the Loschmidt echo. The standard approach considers that the irreversibility
of this phenomenon is the result of an additional term in the backward
Hamiltonian. Here, we use the non-Hermitian formalism, where the time
evolution is non-unitary. Additionally, we compare the characteristic decay
times of this phenomenon with the decoherence ones. We conclude that the
Loschmidt echo and the decoherence can be considered as two aspects of the
same phenomenon, and that there is a mathematical relationship between their
corresponding characteristic times.
\end{abstract}

\section{Introduction}

Gamow vectors have been introduced in the context of quantum unstable
states, also called quantum resonances. Initially, they were used in nuclear
physics to describe radioactive decay \cite{GAM}. As is well known,
experiments have shown that quantum resonances decay exponentially -- at
least for most observable times. Nevertheless, there are deviations of the
exponential decay law \cite{FON} for very short (Zeno effect \cite{MS}) as
well as for very long times (Khalfin effect \cite{KHA}), which are difficult
to be experimentally observed \cite{FGR,RSM}. Noise effects may also
contribute to slight deviations of the exponential law for intermediate
times \cite{FON}. In any case, this exponential law gives a good
approximation, so that resonance states can be very well approximated by
vector states (or wave functions) decaying exponentially with time. These
are the Gamow vectors or Gamow functions. This means that we have either to
generalize time evolution so as to allow for non-unitary time evolutions, or
to extend the Hilbert space to a larger space containing the Gamow vectors.

The exponential decay of Gamow vectors is accomplished if they are defined
as eigenvectors (with complex eigenvalues) of a total Hamiltonian $H=H_0+V$,
where $V$ is a potential responsible for the decay \cite{NA}. Since
Hamiltonians are self-adjoint, this is not possible in the context of
Hilbert space quantum mechanics \cite{BOH,BOHM,BOHM1}. One consistent
solution is to extend the Hilbert space to a rigged Hilbert space, also
called \textit{Gelfand triplet} \cite{GEL} (see \cite{RM} for an alternative
approach with non-Hermitian Hamiltonians). It is built by adding two more
spaces to the Hilbert space $\mathcal{H}$, so that they form a triplet, $%
\Phi\subset\mathcal{H}\subset\Phi^\times$, where $\Phi$ is a locally convex
space, and $\Phi^\times$ is its dual \cite{B,ROB,ANT,MEL,GG,GG1,BEL}. Then,
Gamow vectors are well defined in the larger space $\Phi^\times$. On $%
\Phi^\times$, time evolution can be defined as a non-unitary extension of
the unitary time evolution on the Hilbert space $\mathcal{H}$ \cite%
{BG,CG2004}.

There are several applications where the Gamow vectors (or Gamow states)
play an essential role. The Brussels school used them to develop an
irreversible version of quantum mechanics proposed by Prigogine \cite%
{AP,SATB,ASLT,ADKM,AGPP,PP,AGKPP}. One of its obvious derivations is a
formalism for time asymmetry in quantum mechanics \cite{BOH1,BEU,BOH2},
along with the investigation on a microphysical arrow of time \cite%
{BAK,BAK1,CGL,ACL}. Other applications of Gamow vectors lie on the fields of
quantum decoherence for closed systems \cite{CL,CL1,CG2006}, or even in
non-linear optics \cite{SO,MC,MBGC}. All these applications have in common
the fact that, in the temporal evolution of certain magnitudes, a decreasing
exponential appears that cannot be described by a unitary time evolution.

Recently, we have developed a formalism that relies on the Gamow vectors to
describe the quantum-to-classical transition from the point of view of the
Heisenberg picture \cite%
{Fortin-2016,Losada-2018,Losada-2018a,Fortin-Book-2019}. Its formal
properties were studied from the point of view of dynamical logics \cite%
{Losada-2018a,Fortin-Book-2019}, using an algebraic perspective \cite%
{Losada-2018a}.

The exponential decay also appears in the study of a completely different
phenomenon, namely, the \emph{Loschmidt echo}. A. Peres proposed it in 1984
as a measure of sensitivity and reversibility of quantum evolutions. The
idea is to consider an initial state $|\psi_0\rangle$ and evolve it during a
time interval $\tau$ under the action of a Hamiltonian $H_1$, reaching the
state $|\psi_{\tau}\rangle $. Then, apply a second Hamiltonian $H_2$, during
a time interval of the same duration, in such a way that it reverses the
action of the first Hamiltonian, aiming to recover the initial state. $H_1$
is the Hamiltonian governing the forward evolution and $H_2$ is the
Hamiltonian governing the backwards evolution. Due to imperfections in the
time-reversal or due to an intrinsic quantum irreversibility, the final
state does not always coincide with the initial one. The Loschmidt echo
measures this discrepancy, and it is defined as

\begin{equation}  \label{e:Mt}
M(\tau)=|\langle \psi_0| e^{-i H_2 \tau / \hbar }e^{-i H_1 \tau /
\hbar}|\psi_0 \rangle |^2.
\end{equation}

The above equation means that the system evolves in time under the action of
the Hamiltonian $H_{1}$, and then, the evolution is governed by Hamiltonian $%
H_{2}$. If the time reversal operation is perfectly implemented and the time
evolution is unitary, then $H_{2}=-H_{1}$ and $M(t)=1$. In case that some
imperfections are involved, we can assume that $H_{2}=-H_{1}+\Sigma$, where $%
\Sigma$ is a perturbation of the original Hamiltonian. The perturbation term
is intended to bear information about possible imperfections in the
implementation of the time reversal operation. With these definitions,
equation \eqref{e:Mt} can be written as

\begin{equation}  \label{e:Mt2}
M(\tau)=|\langle \psi_0| e^{i(H_{1}-\Sigma)\tau / \hbar }e^{-i H_1 \tau /
\hbar}|\psi_0 \rangle |^2.
\end{equation}

The time reversal operation has been widely studied from a theoretical point
of view. However, the empirical study of the time reversal problem involves
looking for physical operations for the implementation of the time
inversion, which is not a trivial task. The development of new experimental
techniques has allowed to perform precise experiments in this area. In some
systems, it has been observed that the initial and the final states can
differ to some degree \cite{ Pastawski}. To explain this discrepancy, two
factors must be considered: the unitarity of the time evolution and the
noise during the process.

When only unitary evolutions are considered, the discrepancy can be
attributed to an imperfection in the implementation of the time reversal
operation. In this case, the Hamiltonian which is used to model the temporal
reversion is not exactly the inverse of the initial one. This can be
attributed to the noise introduced by the environment, and it has been
studied in previous works (see for example \cite{Zurek2007}).

In this paper, we propose an alternative description in order to explain the
discrepancy between the initial and final states. We assume that the time
inversion is perfect, but the time evolution is not unitary. To obtain a
non-unitary evolution, we appeal to Gamow vectors, which are adequate for
introducing exponential decays. Using this formalism, we show that the time
evolution generated by Hamiltonians with complex eigenvalues is
irreversible. Moreover, this approach allows us to account for the Loschmidt
echo phenomenon.

\section{Gamow vectors formalism}

In this section, we introduce a model that will be useful to understand our
description of the Loschmidt echo. This model is a variant of the Friedrichs
one \cite{F} with multiple resonances, and it is general enough to describe
a wide range of realistic situations \cite{GP}.

\subsection{The Friedrichs model}

As is well known, \textit{scattering resonances} and \textit{%
quasi-stationary states} are used to describe a scattering process in which
a particle stays in the neighborhood of a center of forces --usually given
by a potential-- a time much longer than it would have remained if the
center of forces were not present \cite{BOH,BEU}. Scattering resonances can
be modeled using Hamiltonian pairs, $\{H_0,H\}$, where $H_0$ is the
unperturbed Hamiltonain and $H=H_0+V$, where $V$ is the potential
responsible for the creation of the quasi-stationary states. The simplest
model for resonances is the Friedrichs model, in which $H_0$ is given by

\begin{equation}  \label{1}
H_0=\omega_0\,|1\rangle\langle 1|+ \int_0^\infty
\omega\,|\omega\rangle\langle\omega|\,d\omega\,.
\end{equation}

Here, $H_0$ has a pure absolutely and non-degenerate continuous spectrum,
which is $\mathbb{R}^+\equiv[0,\infty)$ and an eigenvalue, $\omega_0>0$.
Kets $|\omega\rangle$ are the eigenkets of $H_0$ with eigenvalues $\omega\in 
\mathbb{R}^+$, $H_0|\omega\rangle=\omega\,|\omega\rangle$. As in the case of
plane waves, $|\omega\rangle$ are not normalizable states, which acquire
meaning in suitable extensions of the Hilbert space called rigged Hilbert
spaces or Gelfand triplets, for which there exists an extensive literature
as mentioned in the Introduction. The eigenvector $|1\rangle$, $%
H|1\rangle=\omega_0\,|1\rangle$ belongs to Hilbert space and is normalized
so that $\langle 1|1\rangle=1$. Notice that the eigenvalue $\omega_0$ is
imbedded in the continuous spectrum of $H_0$.

The potential intertwines discrete and continuous spectrum of $H_0$ and it
usually has the form

\begin{equation}  \label{2}
V=\int_0^\infty f(\omega) \,[|\omega\rangle\langle
1|+|1\rangle\langle\omega|]\,d\omega\,,
\end{equation}
where $f(\omega)$ is a real function, usually square integrable, called 
\textit{form factor}. Since bound states and scattering states are
orthogonal to each other, we have that $\langle \omega|1\rangle=\langle
1|\omega\rangle=0$, for all $\omega\in\mathbb{R}^+$. In addition, $%
\langle\omega|\omega^{\prime }\rangle=\delta(\omega-\omega^{\prime })$ \cite%
{GP}.

The total Hamiltonian has the form $H=H_0+\lambda\,V$, where $\lambda$ is a
real coupling constant. The Hamiltonian $H$ has a pure simple absolutely
continuous spectrum, so that

\begin{equation}  \label{3}
H=\int_0^\infty \omega\,|\omega^\pm\rangle\langle\omega^\pm|\,d\omega\,,
\end{equation}
where $H|\omega^\pm\rangle=\omega\,|\omega^\pm\rangle$ are eigenkets for $%
\omega\in\mathbb{R}^+$, the continuous spectrum of $H$. The signs $\pm$
correspond to \textit{in} and \textit{out} states of scattering theory. More
details can be found in \cite{GP}.

This is the simplest possible Friedrichs model. The most straightforward
generalization, is obtained by adding more bound states for $H_0$, so that

\begin{equation}  \label{4}
H_0=\sum_{k=1}^N \omega_k\,|k\rangle\langle k| +\int_0^\infty
\omega\,|\omega\rangle\langle\omega|\,d\omega\,, \qquad \omega_k>0\,.
\end{equation}

\noindent Other generalizations can be found in \cite{GP}, although they do
not play any role in the present discussion.

Thus, while $H_0$ has bound states, at least one, $H$ possesses none. What
has happened with the bound states of $H_0$? To see it, we need a
mathematical definition for resonances. A precise definition based in the
analytic properties of the resolvent $(H-zI)^{-1}$ of $H$, where $z$ is a
complex number, $z\in\mathbb{C}/\mathbb{R}^+$ and $I$ the identity operator,
was given in \cite{RSIV}, page 55. In the case of the simplest Friedrichs
model, in which $H_0$ has only one bound state, it is sufficient to consider
the analytic properties of the \textit{reduced resolvent function} given by

\begin{equation}  \label{5}
\eta(z):= \langle 1|(H-zI)^{-1}|1\rangle\,.
\end{equation}

Under mild conditions on the form factor $f(\omega)$ \cite{EX}, the function 
$\eta(z)$ admits an analytic continuation with a pair of poles at the points 
$z_0=E_R-i\Gamma/2$ and $z_0^*=E_R+i\Gamma/2$, with $\Gamma>0$. In addition,
and following the formulation of resonances in rigged Hilbert spaces \cite%
{BG,CG2004}, there are two non-normalizable vectors $|f^D\rangle$ and $%
|f^G\rangle$, such that

\begin{equation}  \label{6}
H|f^D\rangle = z_0\,|f^D\rangle\,,\qquad H|f^G\rangle =z_0^*\, |f^G\rangle\,.
\end{equation}

\noindent For any $t>0$ or $t<0$, respectively, one has that

\begin{equation}  \label{7}
e^{-itH}\,|f^D\rangle = e^{-iE_R\,t}\,e^{-\Gamma t/2}\,|f^D\rangle\,,\qquad
e^{-itH}\,|f^G\rangle = e^{-iE_R\,t}\,e^{\Gamma t/2}\,|f^G\rangle\,,
\end{equation}

\noindent which shows that $|f^D\rangle$ and $|f^G\rangle$ decay
exponentially as $t\longmapsto +\infty$ and $t\longmapsto-\infty$,
respectively. Thus, the superscripts $D$ and $G$ means decay and grow,
respectively, for times going from $-\infty$ to $\infty$.

Usually, $|f^D\rangle$ and $|f^G\rangle$ receive the names of \textit{%
decaying} and \textit{growing} Gamow vectors, respectively. Under mild
assumptions, $z_0$ is analytic in the coupling constant $\lambda$ and has
the property that $\lim_{\lambda\to 0} z_0=\omega_0$ \cite{EX}.

In correspondence with the two spectral decompositions for $H$ given in %
\eqref{3}, we have these other two:

\begin{equation}  \label{8}
H=z_0\,|f^D\rangle \langle f^G|+ \int_{\gamma^-} z\,|z^-\rangle\langle
z^+|\,dz\,,\qquad H= z_0^*\,|f^G\rangle \langle f^D|+ \int_{\gamma^+}
z\,|z^+\rangle\langle z^-|\,dz\,.
\end{equation}

\begin{figure}[h!]
\centering
\includegraphics[width=10cm]{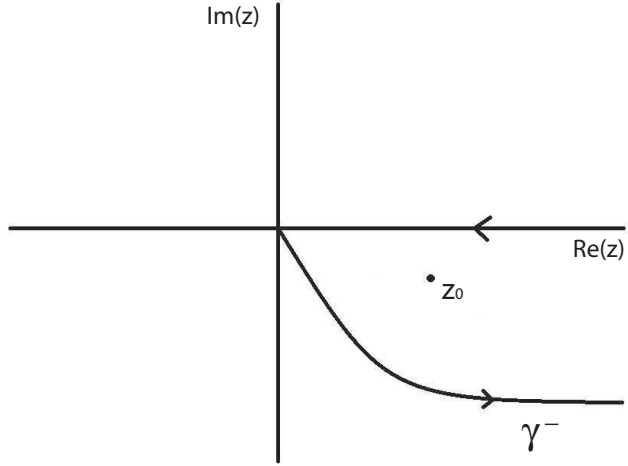}
\caption{ $\protect\gamma^-$ is a curve in the lower half plane, starting at
the origin, that together with the positive semi-axis of the real line
encloses the pole $z_0$.}
\label{fig:figura}
\end{figure}

The meaning of the first spectral decomposition \eqref{8} is the following:
let us assume that the functions $\varphi(\omega) =
\langle\omega|\varphi\rangle$ and $\phi(\omega)=\langle\omega|\phi\rangle$
are analytically continuable to the upper and lower half of the complex
plane, respectively. This, in particular, implies that $\varphi^*(\omega)=
\langle\varphi|\omega\rangle$ admits analytic continuation to the lower half
plane. The value of these functions at the point $z_0$ in the lower half
plane is given by $\varphi^*(z_0)=\langle\varphi|f^D\rangle$ and $%
\phi(z_0)=\langle f^G|\phi\rangle$. The values of $\varphi^*(\omega)$ and $%
\phi(\omega)$ at the complex number $z$ with negative imaginary part are $%
\varphi^*(\omega)=\langle\varphi|z^-\rangle$ and $\phi(z)=\langle
z^+|\phi\rangle$, respectively. Thus, what really has proper meaning is the
value of $\langle \varphi|H|\phi\rangle$. Finally, $\gamma^-$ is a curve in
the lower half plane, starting at the origin, that together with the
positive semi-axis of the real line encloses the pole $z_0$ (see Fig. \ref%
{fig:figura}). Analogously, $\gamma^+$ is a curve in the upper half plane,
starting at the origin, that together with the positive semi-axis of the
real line encloses the pole $z^*_0$. The meaning of the second spectral
decomposition in \eqref{8} is similar, but in the upper half plane.

A Friedrichs model with free Hamiltonian as in \eqref{4} and potential given
by

\begin{equation}  \label{9}
V:=\sum_{k=1}^N \int_{0}^{\infty} f_k(\omega) \,[|\omega\rangle\langle
k|+|k\rangle\langle \omega|]\,d\omega\,,
\end{equation}

\noindent where the $f_k(\omega)$ are $N$ form factor functions (the total
Hamiltonian is $H=H_0+\lambda V$), is expected to give $N$ resonance poles
(although for a given value of $\lambda$, two of these poles may coincide at
the same point, giving a double resonance pole). Thus, for a given value of $%
\lambda$, we have resonance poles at the points $z_1,\dots,z_N$ and their
complex conjugates, with respective decaying and growing Gamow vectors, $%
|f_1^D\rangle,\dots, |f_N^D\rangle$ and $|f_1^G\rangle, \dots, |f_N^G\rangle$%
. Then, the spectral decompositions \eqref{8} generalize to

\begin{eqnarray}  \label{10}
H= \sum_{k=1}^N z_k\, |f_k^D\rangle\langle f_k^G| + \int_{\gamma^-}
z\,|z^-\rangle\langle z^+|\,dz\,, \quad H= \sum_{k=1}^N z^*_k\,
|f_k^G\rangle\langle f_k^D| + \int_{\gamma^+} z\,|z^+\rangle\langle
z^-|\,dz\,.
\end{eqnarray}

The integral terms in \eqref{8} and \eqref{10} are responsible for the
deviations of the exponential decay for very short (Zeno effect \cite{MS})
or very large values of time \cite{KHA}. These effects are observable,
although not easy to be observed \cite{FGR,RSM}. This means that, for the
vast majority of experiments, they are not detected. Due to this reason, it
makes sense to drop the integral term form \eqref{8} and \eqref{10}, at
least within a reasonable range of observations. Also, note that the
spectral decompositions for $H$ either in \eqref{8} or in \eqref{10} are
time reversal of each other, so that they are completely equivalent \cite%
{CGL}. Therefore, one may take as an effective Hamiltonian, valid for the
majority of observations, the following:

\begin{equation}  \label{11}
H= \sum_{k=1}^N z_k\, |f_k^D\rangle\langle f_k^G|\,.
\end{equation}

As in the case $N=1$, we have that $H|f^D_k\rangle=z_k\,|f_k^D\rangle$, for $%
k=1,\dots,N$. Along \eqref{1}, this suggests that

\begin{equation}  \label{12}
\langle f_s^G|f_k^D\rangle = \delta_{s,k}\,,\qquad s,k=1,\dots,N\,.
\end{equation}

\noindent Analogously, $H|f^G_k\rangle=z^*_k\,|f_k^G\rangle$ for all $k$, so
that if we take as effective Hamiltonian the second sum in \eqref{10}, we
have that $\langle f_s^D|f_k^G\rangle=\delta_{s,k}$, for all values of $s$
and $k$. Another property derived from this pseudometrics is that $\langle
f_k^D|f_s^D\rangle =0 = \langle f_k^G|f_s^G\rangle$.

Resonances may be considered as independent of each other, so that the Gamow
vectors $|f_1^D\rangle, \dots, |f_N^D\rangle$ and $|f_1^G\rangle,\dots,
|f_N^G\rangle$ may be assumed to be linearly independent. Then, we may
assume that linear combinations of resonance states may belong either to the 
$N$ dimensional linear space, $\mathcal{H}_N^D$, spanned $|f_1^D\rangle,
\dots, |f_N^D\rangle$ , or the space $\mathcal{H}_N^G$ spanned by $%
|f_1^G\rangle, \dots, |f_N^G\rangle$ or even to the $2N$ dimensional space, $%
\mathcal{H}_{2N}$, spanned by $|f_1^D\rangle, \dots, |f_N^D\rangle$ and $%
|f_1^G\rangle, \dots, |f_N^G\rangle$. Due to \eqref{12}, the spaces $%
\mathcal{H}_N^D$ and $\mathcal{H}_N^G$ may be considered as duals of each
other.

These ideas, along with \eqref{7}, are the basis for the following comment.
Assume that $|\psi\rangle$ and $|\varphi\rangle$ are arbitrary vectors in $%
\mathcal{\ H}_N^D$, so that

\begin{equation}  \label{13}
|\psi\rangle:= \sum_{k=1}^N a_k\,|f_k^D\rangle\,,\qquad |\varphi\rangle:=
\sum_{k=1}^N b_k\,|f_k^D\rangle\,.
\end{equation}

\noindent Let us evolve $|\varphi\rangle$ in time, with the evolution
governed by the total Hamiltonian $H$. The value of this vector after a time 
$t$ has elapsed, is given by

\begin{equation}  \label{14}
|\varphi(t)\rangle = \sum_{k=1}^N b_k\,|f_k^D(t)\rangle\,,
\end{equation}

\noindent where $|f_k^D(t)\rangle =e^{-itH}\,|f_k^D\rangle$ as in \eqref{7}.
Because of the duality between the spaces $\mathcal{H}_N^D$ and $\mathcal{H}%
_N^G$, the bra corresponding to the ket $|\psi\rangle$ is

\begin{equation}  \label{15}
\langle \psi|= \sum_{k=1}^N a^*_k\,\langle f_k^G|\,,
\end{equation}
where the star denotes complex conjugation. The probability amplitude for
the state $|\varphi(t)\rangle$ to be in the state $|\psi\rangle$ at time $t$
is given by

\begin{equation}  \label{16}
A(t)= \langle\psi|\varphi(t)\rangle = \sum_{k,s=1}^N a_s^*\,b_k\,\langle
f_s^G|f_k^D(t)\rangle = \sum_{k,s=1}^N a_s^*\,b_k\, A_{s,k}(t)\,,
\end{equation}

\noindent where,

\begin{equation}  \label{17}
A_{s,k}(t) = \langle f_s^G|e^{-itH}|f_k^D\rangle =e^{-iz_kt}\, \langle
f_s^G|f_k^D\rangle =
e^{-iz_kt}\delta_{s,k}=e^{-iE_kt}\,e^{-t\,\Gamma_k/2}\delta_{s,k}\,,
\end{equation}

\noindent so that

\begin{equation}  \label{18}
A(t)= \sum_{k=1}^N a_k^*\,b_k\,e^{-iE_kt}\,e^{-t\,\Gamma_k/2}\,,
\end{equation}
\noindent where $E_k$ and $-\Gamma_k/2$ are the real and imaginary parts of
the resonance pole $z_k$, respectively.

Instead, we may have considered the vectors $|\psi\rangle$ and $%
|\varphi\rangle$ to be in the space $\mathcal{H}_{2N}$, and follow a similar
reasoning. Formally, the results would have been different, since to %
\eqref{18}, we had to add a similar term. If this were the case,

\begin{equation}  \label{19}
|\psi\rangle:= \sum_{k=1}^N a_k\,|f_k^D\rangle + \sum_{k=1}^N
c_k\,|f^G_k\rangle\,,\qquad |\varphi\rangle:= \sum_{k=1}^N
b_k\,|f_k^D\rangle + \sum_{k=1}^N d_k\,|f^G_k\rangle\,.
\end{equation}

\noindent This construction, although useful for other purposes, may be
incompatible with the formalism of Time Asymmetric Quantum Mechanics (TAQM) 
\cite{BOH1,BEU,BOH2,SIGMA}, since in this formalism, time evolution for $%
|f_k^D\rangle$ is defined \textit{only} for $t\ge 0$, while time evolution
for $|f_k^G\rangle$ is \textit{only} valid for $t\le 0$. Nevertheless, a
construction such that all growing and decaying Gamow vectors evolve
rigorously for all values of time is possible \cite{GL}, although this
construction does not take into account the time asymmetry that is produced
by decaying processes\footnote{%
It is also true that an algebraic formulation using operators on $\mathcal{H}
_{2N}$ is still compatible with TAQM, \cite{Losada-2018}.} Thus, we consider
the discussion on $\mathcal{H}_N^D$ as given above, as the most appropriate
for our purposes.

\subsection{Creation and annihilation operators}

In this section we study the Friedrichs model in terms of the creation and
annihilation operators. Its construction has been proposed, for instance, in 
\cite{AGKPP}. The idea is the following: Let us assume that $a$ and $%
a^\dagger$ are the annihilation and creation of $|1\rangle$ in \eqref{1},
respectively, and that $b_\omega$ and $b^\dagger_\omega$ play the same role
in relation to $|\omega\rangle$. Then, the Friedrichs free Hamiltonian $H_0$
written in terms of these operators is a contribution of two terms $%
H_0=H_A+H_R$, where \cite{AGKPP},

\begin{eqnarray}  \label{41}
H_A:= \omega_0\, a^\dagger\,a\,,\qquad H_R:= \int_0^\infty
d\omega\,\omega\,b_\omega^\dagger\,b_\omega\,,
\end{eqnarray}
and the interaction is now written as

\begin{equation}  \label{42}
H_I= \lambda \int_0^\infty d\omega\,f(\omega)
(a^\dagger\,b_\omega+a\,b^\dagger_\omega)\,,
\end{equation}
where $\lambda$ is the coupling constant and $f(\omega)$ the form factor.
Again, the total Hamiltonian is $H=H_0+H_I=H_A+H_R+H_I$. The function

\begin{equation}  \label{43}
\eta(z):= \omega_0 - z -\int_0^\infty d\omega\; \frac{\lambda^2\,f^2(\omega) 
}{\omega-z}\,,
\end{equation}
is analytic with a branch cut on the positive semi-axis $[0,\infty)$. We
denote its limits on this semi-axis from above to below and from below to
above as $\eta^+(\omega):=\eta(\omega+i0)$ and $\eta^-(\omega):=\eta(
\omega-i0)$, respectively. Then, a ``diagonalization'' of the total
Hamiltonian $H$ is given by \cite{AGKPP}

\begin{equation}  \label{44}
H= z_0\, (A^\dagger)_{\mathrm{in}}\,A_{\mathrm{out}} + \int_0^\infty
d\omega\,\omega \; \frac{\eta^+(\omega)}{\eta^-(\omega)}\; (\tilde
B^\dagger_\omega)_{\mathrm{in}}\, (\tilde B_\omega)_{\mathrm{out}}\,.
\end{equation}
Here, $(A^\dagger)_{\mathrm{in}}$ and $A_{\mathrm{out}}$ are the creation
and annihilation of the Gamow state from a vacuum $|0\rangle$, so that

\begin{equation}  \label{45}
|f^D\rangle = (A^\dagger)_{\mathrm{in}}\,|0\rangle\,, \quad |0\rangle= A_{ 
\mathrm{out}} |f^D\rangle\,,\quad A_{\mathrm{out}}|0\rangle=0\,, \quad [A_{ 
\mathrm{out}}, A^\dagger_{\mathrm{in}}]=1\,,
\end{equation}
while $(\tilde B^\dagger_\omega)_{\mathrm{in}}$ and $(\tilde B_\omega)_{ 
\mathrm{out}}$ are creation and annihilation operators of states in the
continuum, which were defined in \cite{AGKPP}. The latter do not have here
any interest for us, as we are solely interested in the resonance behavior.
The subscripts ``in'' and ``out'' come from scattering theory and do not
play any role in our discussion here, so that we will omit them. Thus, the
effective Hamiltonian, which concerns only to the presence of resonances is
given by $H=z_0\,A^\dagger\,A$ in our notation.

Now, rename $|f_1^D\rangle:=|f^D\rangle$ and define $|f_n^D\rangle:=
(A^\dagger)^n\,|f_1^D\rangle$. If we denote by $N:= A^\dagger\,A$, it
readily comes that

\begin{equation}  \label{46}
N|f_n^D\rangle =n \,|f_n^D\rangle\,,\qquad H|f_n^D\rangle
=nz_0\,|f_n^D\rangle\,,
\end{equation}
which formally yields

\begin{equation}  \label{47}
e^{-itH}\,|f_n^D\rangle = e^{-itnz_0}\,|f_n^D\rangle\,.
\end{equation}

In the next section, we are going to apply the formalism of Gamow vectors to
the Loschmidt echo.

\section{Loschmidt echo and Gamow vectors}

We consider a quantum system in an initial state $|\alpha (0)\rangle $. Let
us suppose that this initial state is evolved during a time interval $\tau $%
, and then it is evolved backwards during the same interval $\tau$. We
denote the final state as $|\alpha (2\tau )\rangle $. At this point, we can
compare the initial state $|\alpha (0)\rangle $ with the final state $%
|\alpha (2\tau )\rangle$. When the time-reversal procedure is perfectly
implemented and the time-evolution is unitary, both states are exactly the
same. However, when the time-reversal implementation is imperfect, there
exist some degree of discrepancy between the states. One way to quantify
this discrepancy is using the Loschmidt echo. If the time-evolution of the
quantum system is given by the operator $U_{1}(\tau)=e^{-\frac{i}{\hbar}%
H_{1}\tau}$, we define the decay rate by

\begin{equation}  \label{M}
M(\tau ):=|\langle \alpha (0)|U_{2}(\tau )\,U_{1}(\tau )|\alpha (0)\rangle
|^{2}\,.
\end{equation}

\noindent where $U_{2}(\tau)=e^{\frac{i}{\hbar}H_{2}\tau}$, and $H_{2}$ is a
perturbation of $H_{1}$.

It is easy to see that $M(\tau )=1$, when there is no discrepancy, and $%
M(\tau )=0$, when the discrepancy is maximal. In particular, $M(\tau )$ is
maximal when the time evolution is unitary and the time reversal is perfect.
In the usual approach, it is assumed that the time evolution is unitary, but
the time reversal is imperfect, i.e., the backward Hamiltonian is not
exactly opposite to the forward Hamiltonian. In this work, we adopt an
alternative approach. We consider a perfectly implemented time reversal, but
a non-unitary time evolution.

In order to illustrate our approach, let us discuss how the time reversal is
implemented in laboratory realizations. Consider the case of a crystal under
the action of an external magnetic field oriented in the $\hat{z}$ direction
(as discussed in \cite{Pastawski-1998}). After applying $U_{1}$ during a
time interval $\tau$, the orientation of the external magnetic field is
rotated and a new Hamiltonian $H_2$ is obtained. The action of $H_2$ during
a time interval $\tau$ is formally equivalent to the time reversion.

In the standard approach, the new Hamiltonian $H_2$ is represented as a
perturbation of the opposite initial Hamiltonian $H_1$. In our alternative
description, we assume that evolutions are non-unitary. This is the
mathematical expression of the assumption that the system has an intrinsic
irreversibility which is not caused by the environment. The irreversibility
is formally represented by the decaying exponential factor related to the
time evolution of a Gamow vector.

Let us describe how the second Hamiltonian $H_2$ is modeled under these
assumptions. In the crystal example, the magnetic field is changed from $%
\hat{z}$ to $-\hat{z}$. After this transformation, the signs of the real
parts of the eigenvalues change, but the resonances remain the same, because
they are related to internal degrees of freedom, and have nothing to do with
the orientation of the system respect to the laboratory. Thus, the
Hamiltonian $H_2$ changes its complex eingenvalues as follows: $z = \omega-%
\frac{i}{2}\Gamma \longrightarrow -z^* = -\omega-\frac{i}{2}\Gamma$.
Physically, this means that reversing the physical process do not change the
decay factor of the dynamical law. Our description of what happens in the
laboratory allows to explain why the initial state is not recovered. The
decaying behavior of the Gamow vectors explains the Lodschmit echo.
Therefore, the forward and backward evolution operators, $U_1(\tau)$ and $%
U_2(\tau)$ respectively, have the following form: 
\begin{align}  \label{63}
U_1(\tau)&= \sum_{n=0}^N e^{-\frac{i \tau}{\hbar} z_n}\,
|f_n^D\rangle\langle f_n^G| \,. \\
U_2(\tau)&= \sum_{n=0}^N e^{\frac{i \tau}{\hbar} z^*_n}\,
|f_n^D\rangle\langle f_n^G| \,.
\end{align}
Notice that, for all Gamow states $\{|f_n^D\rangle\}$, we have

\begin{equation}  \label{64}
U_1(\tau)|f_n^D\rangle = e^{-\frac{i \tau}{\hbar} z_n}\,|f_n^D\rangle \qquad 
\mathrm{\ and} \qquad U_2(\tau) |f_n^D\rangle = e^{\frac{i \tau}{\hbar}
z^*_n}\, |f_n^D\rangle\,,
\end{equation}
$U_2(\tau)U_1(\tau)$ is well defined, and it is different from the identity.
Applying this time evolution in equation \eqref{M}, we obtain 
\begin{eqnarray}  \label{m sin condiciones iniciales}
\langle \alpha(0)|U_2(\tau )U_1(\tau )|\alpha(0)\rangle &=&\langle
\alpha(0)|\sum_{n=0}^{N}e^{\frac{i\tau }{\hbar }z_{n}^{\ast
}}\,|f_{n}^{D}\rangle \langle f_{n}^{G}|\sum_{m=0}^{N}e^{-\frac{i\tau }{
\hbar }z_{m}}\,|f_{m}^{D}\rangle \langle f_{m}^{G}|\alpha (0)\rangle = 
\notag  \label{65} \\[2ex]
&=&\sum_{n=0}^{N}e^{\frac{i\tau }{\hbar }(z_{n}^{\ast }-z_{n})}\langle
\alpha (0)|f_{n}^{D}\rangle \langle f_{n}^{G}|\alpha (0)\rangle=  \notag \\
&=&\sum_{n=0}^{N}e^{-\tau \gamma _{n}/\hbar }\langle \alpha
(0)|f_{n}^{D}\rangle \langle f_{n}^{G}|\alpha (0)\rangle \,.
\end{eqnarray}

Let us suppose that the initial state is a superposition of the Gamow
vectors $|f_{n}^{D}\rangle$ 
\begin{equation}
|\alpha(0)\rangle= \sum_{n=0}^N a_{n} \;|f_n^D\rangle,
\end{equation}
then 
\begin{equation}
\langle \alpha(0)|= \sum_{n=0}^N a^*_{n} \; \langle f_n^G |.
\end{equation}
Now, using the relations \eqref{12}, we obtain 
\begin{eqnarray}  \label{suma polos}
\langle \alpha(0)|U_2(\tau )U_1(\tau )|\alpha(0)\rangle =
\sum_{n=0}^{N}e^{-\tau \gamma _{n}/\hbar }|a_n|^2.
\end{eqnarray}
This result shows that the Loschmidt echo decays with time, and its decay
depends on the initial conditions $a_n$ and the characteristic times $%
\gamma_n^{-1}$.

The decay of the Loschmidt echo has different regimes, such as parabolic,
exponential, gaussian among others \cite{paper1, paper2, paper3, paper4,
paper5, paper6, paper7}. The most simple case of equation \eqref{suma polos}
is when there is only one characteristic time, or when the initial
conditions are such that only one mode is activated. In these cases, it is
obtained a pure exponential decay. In this way, Gamow vectors can be used to
model experimental situations. In the general case, the combination of
exponential functions with different characteristic times leads to more
diverse time evolutions.

\section{Decoherence and Loschmidt echo}

So far we have seen how the Gamow vectors can be used to describe the
Loschmidt echo. We saw that the type of decay of the quantity $M(\tau)$ is
determined by the complex part of the Hamiltonian eigenstates and the
initial conditions. In what follows we will show that there is a close
relationship between the Loschmidt echo and the quantum decoherence
phenomenon. This connection is because the decoherence time is also related
with the imaginary part of the Hamiltonian eigenstates and the initial
conditions.

To study the decoherence process in the Friedrich model, it is necessary to
introduce the quasi coherent states, since they form the so-called
privileged basis. Quasi-coherent states have been discussed in \cite%
{SKA,AAG,POP}.

\subsection{Initial conditions}

For systems having an infinite number of resonances, it is possible in
principle to define coherent resonance states. This states should have the
following form for each complex number $z\in\mathbb{C}$:

\begin{equation}  \label{20}
|z^D\rangle =e^{-|z|^2/2}\sum_{k=0}^\infty \frac{z^n}{\sqrt{n!}}
\,|f_k^D\rangle\,.
\end{equation}

\noindent These coherent states should be eigenvectors with eigenvalue $z$
of a annihilation operator $B^-$, satisfying $B^-|f_k^D\rangle=k|f_{k-1}^D%
\rangle$, for all $k\in \mathbb{N}$, where $|f_0^D\rangle=0$, so that $%
B^-|z^D\rangle=z\,|z^D\rangle$. This can be implemented for some exactly
solvable models for resonances, like the high barrier P\"oschl-Teller one
dimensional model \cite{CGKN, CG1}.

Unfortunately, up to our knowledge, nobody has constructed a solvable
Friedrichs model with an infinite number of resonance poles. Let us assume
that we have $N+1$ resonances with decaying Gamow vectors $|f_0^D\rangle,
\dots,|f_N^D\rangle$. By definition, a \textit{quasi coherent state} at time 
$t=0$ is a vector of the form:

\begin{equation}  \label{21}
|\alpha_1(0)\rangle:= \left(\sum_{k=0}^N \frac{|\alpha_1(0)|^{2k}}{k!}
\right)^{-1/2} \sum_{n=0}^N \frac{(\alpha_1(0))^n}{\sqrt{n!}}
\;|f_n^D\rangle\,,
\end{equation}
where $\alpha_1(0)$ is a given complex number. Note that for $N$ going to
infinity \eqref{21} acquires the form \eqref{20}, hence the name of
quasi-coherent state. Analogously, for any other complex number $\alpha_2(0)$%
, we have

\begin{equation}  \label{22}
|\alpha_2(0)\rangle:= \left(\sum_{k=0}^N \frac{|\alpha_2(0)|^{2k}}{k!}
\right)^{-1/2} \sum_{n=0}^N \frac{(\alpha_2(0))^n}{\sqrt{n!}}
\;|f_n^D\rangle\,.
\end{equation}

\noindent Notice that, in a rough approximation, we may consider the pair of
vectors $\{|\alpha_1(0)\rangle,|\alpha_2(0)\rangle\}$ as the \textit{Moving
Preferred Basis} in the sense of \cite{CF}. The approximation is rough
because we have taken $t=0$, therefore it is not properly a moving basis. It
is certainly efficient in the range of extremely short decoherence times $%
t_D $ compared to relaxation time $t_R$, i.e., $t_D<<t_R$. In this sense,
the choice $t_D=0$ is a workable approximation. Let us consider an arbitrary
normalized linear combination of these two vectors; $|\Phi(0)\rangle :=
a|\alpha_1(0)\rangle+ b|\alpha_2(0)\rangle$. Its corresponding density
matrix is given by 
\begin{align}  \label{23}
\rho_0=& |\Phi(0)\rangle\langle\Phi(0)| =  \notag \\
=&|a|^2\,|\alpha_1(0)\rangle\langle\alpha_1(0)| +
ab^*\,|\alpha_1(0)\rangle\langle\alpha_2(0)|
+a^*b\,|\alpha_2(0)\rangle\langle\alpha_1(0)| +|b|^2\,
|\alpha_2(0)\rangle\langle\alpha_2(0)|\,.
\end{align}
Note that, if $i=1,2$,

\begin{equation}  \label{24}
\langle\alpha_i(0)|= \left(\sum_{k=0}^N \frac{|\alpha_i(0)|^{2k}}{k!}
\right)^{-1/2} \sum_{n=0}^N \frac{(\alpha^*_i(0))^n}{\sqrt{n!}}\; \langle
f_n^G|\,.
\end{equation}
After a time $t>0$, the state becomes

\begin{align}  \label{25}
\rho(t)= &|\Phi(t)\rangle\langle\Phi(t)| =  \notag \\
=& |a|^2\,|\alpha_1(t)\rangle\langle\alpha_1(t)| +
ab^*\,|\alpha_1(t)\rangle\langle\alpha_2(t)|
+a^*b\,|\alpha_2(t)\rangle\langle\alpha_1(t)| +|b|^2\,
|\alpha_2(t)\rangle\langle\alpha_2(t)|\,.
\end{align}
The expression in the second row in \eqref{25} contains four terms. The sum
of the second and third terms is called the non-diagonal part that we intend
to analyze next.

\subsection{Decoherence}

Let us denote the sum of the second and third term in \eqref{25} as $\rho^{%
\mathrm{ND}}(t)$. Under the condition of macroscopicity, which is introduced
below, the non-diagonal terms in \eqref{25} approximately satisfy the
following relation:

\begin{equation}  \label{26}
\rho^{\mathrm{ND}}(t) = \rho^{\mathrm{ND}}_{12}(t)
\,|\alpha_1(0)\rangle\langle\alpha_2(0)| + \rho^{\mathrm{ND}}_{21}(t) \,
|\alpha_2(0)\rangle\langle\alpha_1(0)|\,.
\end{equation}
This is not true in the general case as one may check by a direct
calculation on $|\alpha_1(t)\rangle\langle\alpha_2(t)| = e^{-itH}\,
|\alpha_1(0)\rangle\langle\alpha_2(0)| \,e^{itH}$.

In order to show \eqref{26}, we should make the hypothesis of the
macroscopicity condition, which means that the peaks of both approximate
Gaussians are far from each other. This means that $|\alpha_1^*(0)-%
\alpha_2(0)|$ is very large (we will consider $\alpha_1(0)$ as a real
value). Notice that, under this hypothesis, the states of the preferred
basis $\{|\alpha_1(0)\rangle,|\alpha_2(0)\rangle\}$ are approximately
orthogonal or quasi-orthogonal, which means that

\begin{equation}  \label{27}
\langle
\alpha_1(0)|\alpha_2(0)\rangle=\langle\alpha_2(0)|\alpha_1(0)\rangle^*
\approx 0\,.
\end{equation}

\noindent To show \eqref{27}, we first note that after \eqref{22} and %
\eqref{24} we have that

\begin{equation}  \label{28}
\langle \alpha_1(0)|\alpha_2(0)\rangle= \left(\sum_{k=0}^N \frac{%
|\alpha_1(0)|^{2k}}{k!} \right)^{-1/2} \left(\sum_{k=0}^N \frac{%
|\alpha_2(0)|^{2k}}{k!} \right)^{-1/2} \sum_{n=0}^N \frac{%
(\alpha_1^*(0)\,\alpha_2(0))^n}{n!}\,.
\end{equation}

\noindent Then, let us use the Cauchy product\footnote{%
The Cauchy product states that 
\begin{equation*}
\left(\sum_{n=0}^N a_n\right) \left( \sum_{n=0}^N b_n \right) = \sum_{n=0}^N
c_n\,, \qquad \mathrm{with} \qquad c_n = \sum_{k=0}^n a_k\,b_{n-k}\,.
\end{equation*}%
} and the binomial formula on the two first factors on the right hand side
of \eqref{28}, so as to obtain:

\begin{equation}  \label{29}
\langle \alpha_1(0)|\alpha_2(0)\rangle= \left(\sum_{k=0}^N \frac{%
(|\alpha_1(0)|^2 + |\alpha_2(0)|^2)^k}{k!} \right)^{-1/2} \sum_{n=0}^N \frac{%
(\alpha_1^*(0)\,\alpha_2(0))^n}{n!}\,.
\end{equation}

\noindent Repeating the procedure, we obtain

\begin{eqnarray}  \label{30}
\langle \alpha_1(0)|\alpha_2(0)\rangle= \sum_{n=0}^N \frac1{n!} \left(-\frac
12 (|\alpha_1(0)|^2+|\alpha_2(0)|^2-2\alpha_1^*(0)\alpha_2(0)) \right)^n 
\notag \\
[2ex] =\sum_{n=0}^N \frac1{n!} \left( -\frac{(\alpha_1^*(0)-\alpha_2(0))^2}{2%
} \right)^n\,.
\end{eqnarray}

\noindent Clearly after \eqref{30}, for $|\alpha_1^*(0)-\alpha_2(0)|$ very
large, we obtain \eqref{27}. For this case, let us define

\begin{equation}  \label{31}
\rho_{12}^{\mathrm{ND}}(t):=
\langle\alpha_1(0)|\rho(t)|\alpha_2(0)\rangle\,,\quad \rho_{21}^{\mathrm{ND}%
}(t):= \langle\alpha_2(0)|\rho(t)|\alpha_1(0)\rangle\,,
\end{equation}
where $\rho(t)$ has been introduced in \eqref{25}, from which we have that

\begin{eqnarray}  \label{32}
\rho_{12}^{\mathrm{ND}}(t)= ab^*\,\langle
\alpha_1(0)|\alpha_1(t)\rangle\langle \alpha_2(t)|\alpha_2(0)\rangle + a^*b
\, \langle \alpha_1(0)|\alpha_2(t)\rangle\langle
\alpha_1(t)|\alpha_2(0)\rangle\,,  \notag \\
[2ex] \rho_{21}^{\mathrm{ND}}(t)= ab^*\,\langle
\alpha_2(0)|\alpha_1(t)\rangle\langle \alpha_2(t)|\alpha_1(0)\rangle + a^*b
\, \langle \alpha_2(0)|\alpha_2(t)\rangle\langle
\alpha_1(t)|\alpha_1(0)\rangle\,.
\end{eqnarray}

\noindent In order to compute \eqref{32}, let us follow the notation used in %
\eqref{13}, \eqref{14} and \eqref{15}. In what follows, we will assume that
the numbers $\alpha_1(0)$ and $\alpha_2(0)$ are real. If we choose $%
|\psi\rangle:= |\alpha_1(0)\rangle$ and $|\varphi(t)\rangle:=
|\alpha_1(t)\rangle$, and take into account \eqref{21} and \eqref{24}, we
have that the coefficients $a_k$ and $b_k$ in \eqref{13} and \eqref{14} have
the following form:

\begin{equation}  \label{33}
a_{k}= \frac{\left( \alpha_{1}(0)\right) ^{k}}{\sqrt{k!}}\, e^{-\frac 12 \,|
\alpha_{1}(0)|^2}\,,\qquad b_k= \frac{\left( \alpha_{1}(0)\right) ^{k}}{%
\sqrt{k!}}\, e^{-\frac 12\,| \alpha_{1}(0)|^2}\,,
\end{equation}
so that,

\begin{equation}  \label{34}
\left\langle \alpha_{1}(0)|\alpha_{1}(t)\right\rangle =e^{-\left\vert
\alpha_{1}(0)\right\vert ^{2}}\sum_{n=0}^{N}\frac{\left( \left\vert
\alpha_{1}(0)\right\vert ^{2}\right) ^{n}}{n!}e^{-iz_{n}t}\,.
\end{equation}

\noindent Then, we take $|\psi\rangle:= |\alpha_1(0)\rangle$ and $%
|\varphi(t)\rangle:= |\alpha_2(t)\rangle$, which gives after \eqref{22} and %
\eqref{24} that

\begin{equation}  \label{35}
a_{n}= \frac{\left\vert \alpha_{1}(0)\right\vert ^{n}}{\sqrt{n!}}\,e^{-\frac
12\,|\alpha_{1}(0)|^2}\,,\qquad b_{n}= \frac{\left\vert
\alpha_{2}(0)\right\vert ^{n}}{\sqrt{n!}}\,e^{-\frac
12\,|\alpha_{2}(0)|^2}\,,
\end{equation}
\noindent so that

\begin{equation}  \label{36}
\left\langle \alpha_{1}(0)|\alpha_{2}(t)\right\rangle =e^{-\frac 12\,\{|
\alpha_{1}(0)| ^{2}+\left\vert \alpha_{2}(0)\right\vert ^{2}\}
}\sum_{n=0}^{N}\frac{\left( \left\vert \alpha_{1}(0)\right\vert \left\vert
\alpha_{2}(0)\right\vert \right) ^{n}}{n!}e^{-iz_{n}t}\,.
\end{equation}

\noindent In the third place, we choose $|\psi\rangle= |\alpha_2(0)\rangle$
and $|\varphi(t)\rangle= |\alpha_1(t)\rangle$, which give

\begin{equation}  \label{37}
a_{n}= \frac{\left( \alpha_{2}(0)\right) ^{n}}{\sqrt{n!}}\,e^{-\frac
12\,|\alpha_{2}(0)|^2}\,,\qquad b_{n}=\frac{\left( \alpha_{1}(0)\right) ^{n}%
}{\sqrt{n!}}\, e^{-\frac 12\,|\alpha_1(0)|^2}\,,
\end{equation}
so that

\begin{equation}  \label{38}
\left\langle \alpha_{2}(0)|\alpha_{1}(t)\right\rangle = e^{-\frac 12\,\{|
\alpha_{1}(0)| ^{2}+\left\vert \alpha_{2}(0)\right\vert ^{2}\} }
\sum_{n=0}^{N}\frac{\left( \left\vert \alpha_{1}(0)\right\vert \left\vert
\alpha_{2}(0)\right\vert \right) ^{n}}{n!}e^{-iz_{n}t}
\end{equation}

\noindent Finally, the choice $|\psi\rangle:= |\alpha_2(0)\rangle$ and $%
|\varphi(t)\rangle:= |\alpha_2(t)\rangle$ gives

\begin{equation}  \label{39}
a_n= \frac{\left( \alpha_{2}(0)\right) ^{n}}{\sqrt{n!}}\,e^{-\frac
12\,|\alpha_{2}(0)|^2}\,,\qquad b_n= \frac{\left( \alpha_{2}(0)\right) ^{n}}{%
\sqrt{n!}}\,e^{-\frac 12\,|\alpha_{2}(0)|^2}\,,
\end{equation}
so that,

\begin{equation}  \label{40}
\left\langle \alpha_{2}(0)|\alpha_{2}(t)\right\rangle =e^{-\left\vert
\alpha_{2}(0)\right\vert ^{2}}\sum_{n=0}^{N}\frac{\left( \left\vert
\alpha_{2}(0)\right\vert ^{2}\right) ^{n}}{n!} \, e^{-iz_{n}t}\,.
\end{equation}

So far, we have presented our discussion on the quasi-coherent states based
in the properties of the Friedrichs model. In the next sections, we will
present another approach, in such a way that the above discussion can be
applied to decoherence and the Loschmidt echo.

In this case, we may redefine \eqref{21} and \eqref{22}, and then

\begin{equation}  \label{48}
|\alpha_i(0)\rangle = e^{-\frac 12\,|\alpha_i(0)|^2} \sum_{n=0}^\infty \frac{%
(\alpha_i(0))^n}{\sqrt{n!}}\,|f_n^D\rangle\,, \qquad i=1,2\,.
\end{equation}

Following the discussion in \cite{CF}, we repeat the procedure outlined in
Section 3.1 to the present situation. Now, $z_n=nz_0$ and the sum over $n$
goes to infinite. As initial values, we choose 
\begin{equation}  \label{49}
\alpha_1(0)=0\,,\qquad \alpha_2(0) \neq 0.
\end{equation}
The macroscopicity condition now means that $\alpha_2(0)>>1$. Then,
equations \eqref{34}, \eqref{36}, \eqref{38} and \eqref{40} now give,
respectively,

\begin{equation}  \label{50}
\langle\alpha_1(0)|\alpha_1(t)\rangle =1\,, \quad
\langle\alpha_1(0)|\alpha_1(t)\rangle = e^{-\frac12\,|\alpha_2(0)|^2}
\approx 0\,, \quad \langle\alpha_2(0)|\alpha_1(t)\rangle =
e^{-\frac12\,|\alpha_2(0)|^2} \approx 0\,,
\end{equation}

\begin{equation}  \label{51}
\langle\alpha_2(0)|\alpha_2(t)\rangle = \exp \left\{
-|\alpha_2(0)|^2\right\}\, e^{-iz_0t} \,.
\end{equation}
Consequently, the non-diagonal coefficients defined in \eqref{26}, and
explicitly given in its general form in \eqref{32}, take in this case the
approximate following values: 
\begin{equation}  \label{52}
\rho_{12}^{\mathrm{ND}}(t) \approx ab^*\, \exp \left\{
-|\alpha_2(0)|^2\right\}\, e^{-iz_0t}\,,
\end{equation}
and

\begin{equation}  \label{53}
\rho_{21}^{\mathrm{ND}} \approx a^*b\, \exp \left\{
-|\alpha_2(0)|^2\right\}\, e^{-iz_0t}\,.
\end{equation}
We see that as $t\longmapsto\infty$, the non-diagonal terms are of the order
of $\exp\{ -|\alpha_2(0)|^2 \}\approx 0$, which shows decoherence in the
preferred basis.

This formalism may be extended to a Friedrichs model with $N+1$ bound states
for $H_0$ and, hence, with $N+1$ resonance poles, $z_0,\dots,z_N$. In this
case, equations \eqref{52} and \eqref{53} take the following forms,
respectively,

\begin{equation}  \label{54}
\rho_{12}^{\mathrm{ND}}(t) \approx ab^*\, \exp \left\{
-|\alpha_2(0)|^2\right\} \sum_{n=0}^N \frac{(|\alpha_2(0)|^2)^n}{n!}
\,e^{-\frac i\hbar z_n t}
\end{equation}
and

\begin{equation}  \label{55}
\rho_{12}^{\mathrm{ND}}(t) \approx a^*b\, \exp \left\{
-|\alpha_2(0)|^2\right\} \sum_{n=0}^N \frac{(|\alpha_2(0)|^2)^n}{n!}
\,e^{-\frac i\hbar z_n t}\,.
\end{equation}
These results show that the non-diagonal elements of the density matrix go
to cero when time goes to infinite. This process is called decoherence. In
previous works \cite{OMN,GDS,ZHP}, more examples of this phenomenon have
been studied using non-hermitian Hamiltonians.

As is known, the decoherence is an irreversible process \cite{OMN1},
therefore there is some kind of irreversibility in this model. This
motivates us to study more aspects of the irreversibility of this model. In
particular, in this work we are interested in the Loschmidt echo phenomenon.

\subsection{Loschmidt echo}

As we have seen in equation \eqref{m sin condiciones iniciales}, $M(\tau)$
can be expressed as 
\begin{eqnarray}
M(\tau) =\sum_{n=0}^{N}e^{-\tau \gamma _{n}/\hbar }\langle \alpha
_{2}(0)|f_{n}^{D}\rangle \langle f_{n}^{G}|\alpha _{2}(0)\rangle \,.
\end{eqnarray}
Now, using the identity \eqref{22} and relations \eqref{12}, we have that 
\begin{eqnarray}  \label{66}
\langle f_n^G|\alpha_2(0)\rangle = \left(\sum_{k=0}^N \frac{
|\alpha_2(0)|^{2k}}{k!} \right)^{-1/2} \frac{(\alpha_2(0))^n}{\sqrt{n!}}\,.
\end{eqnarray}
Analogously, using \eqref{24}, one has that 
\begin{equation}  \label{67}
\langle \alpha_2(0)| f_n^D\rangle = \left(\sum_{k=0}^N \frac{
|\alpha_2(0)|^{2k}}{k!} \right)^{-1/2} \frac{(\alpha_2^*(0))^n}{\sqrt{n!}}\,,
\end{equation}
so that 
\begin{equation*}
\langle \alpha _{2}(0)|M(\tau)|\alpha _{2}(0)\rangle =A\sum_{n=0}^{N}\frac{%
|\alpha _{2}(0)|^{2n}}{n!}e^{-\tau \gamma _{n}/\hbar }
\end{equation*}
with 
\begin{equation}
A=\left( \sum_{k=0}^{N}\frac{|\alpha _{2}(0)|^{2k}}{k!}\right) ^{-1}\approx
\left( \sum_{k=0}^{\infty }\frac{|\alpha _{2}(0)|^{2k}}{k!}\right)
^{-1}=\exp \{-|\alpha _{2}(0)|^{2}\}\,  \label{69}
\end{equation}
and the latter approximation makes sense for $N$ large, so that 
\begin{equation}  \label{m con condiciones iniciales}
\langle \alpha _{2}(0)|M(\tau)|\alpha _{2}(0)\rangle \approx e^{-|\alpha
_{2}(0)|^{2}}\sum_{n=0}^{N}\frac{|\alpha _{2}(0)|^{2n}}{ n!}e^{-\tau \gamma
_{n}/\hbar }
\end{equation}

Comparing equation \eqref{m con condiciones iniciales} with equations %
\eqref{54} and \eqref{55}, it is observed that, in both cases, the decay
rate is related to the imaginary part of the Hamiltonian eigenvalues. In the
first case, the decay is accompanied by an oscillation, while in the second
case it is not. In both cases the characteristic times of the exponential
decays are the same.

The result shows that there is an intimate relationship between both
phenomena. This connection is not surprising since the Loschmidt echo and
quantum decoherence can be considered as two expressions of the
irreversibility of quantum systems. However, it is of significant importance
since it links two phenomena that have been studied separately.

\section{Conclusions}

The Gamow vectors formalism has been developed in order to give a
mathematical description of quantum decay phenomena, such as radioactive
phenomena, scattering or decoherence. In this paper we extend the scope of
this formalism. We use the Gamow vectors to describe the Loschmidt echo.

The irreversible character of the Loschmidt echo can be studied in two
manners. One approach considers that the time evolution is perfectly
unitary, but there are imperfections in the implementation of the time
reversal operation. The backward and forward evolution process are governed
by two slightly different Hamiltonians. This can be attributed to the noise
introduced by the environment. The other approach is to assume that the time
evolution operator is not unitary, which introduces an intrinsic
irreversibility of the time evolution. This kind of evolutions are well
represented by the Gamow vector formalism.

In this paper, we adopted the second approach: we used Gamow vectors to
describe the Loschmidt echo phenomenon. We showed that the characteristic
time of this process is directly associated with the resonances of the
analytical continuation of the Hamiltonian. Moreover, we proved that the
type of decay obtained is related to the number of resonances of the system
and their relative weight in the initial state.

Additionally, we compared this phenomenon with another well known one: the
decoherence process. In this case, the resonances are related to the
decoherence and relaxation times. We considered a toy model based on the
Friedrichs model. We showed that choosing as the initial state a
superposition of two quasi-coherent states, and taking into account the
macroscopicity condition, decoherence and Loschmidt echo can be described
simultaneously. Moreover, the characteristic time of the Loschmidt echo is
related with the decoherence time, since both times are determined by the
same resonances. Therefore, we have concluded that, under our hypothesis,
the Loschmidt echo and decoherence, can be considered as two aspects of the
same phenomenon, and that there is a mathematical relation between their
corresponding characteristic times.

\section*{Acknowledgements}

This research was partially supported by grants of FONCYT, CONICET,
Universidad Austral, Universidad de Buenos Aires; and the Junta de Castilla
y Le\'on and FEDER projects (BU229P18 and VA137G18).

F.H. is partially supported by ``RASSR40341: Per un'estensione semantica
della Logica Computazionale Quantistica - Impatto teorico e ricadute
implementative".

\end{document}